\documentclass[]{article}
\usepackage{icrctc07}

\newcommand{\HESSONE}{HESS~J1718$-$385}
\newcommand{\PSRONE}{PSR~J1718$-$3825}
\newcommand{\HMS}[3]{$#1^{\mathrm{h}}#2^{\mathrm{m}}#3^{\mathrm{s}}$}

\title{Discovery of the candidate pulsar wind nebula HESS J1718-385 in very-high-energy gamma-rays}
\shorttitle{Discovery of the candidate pulsar wind nebula HESS J1718-385} 
\authors{S.~Carrigan$^{1}$, Y.A.~Gallant$^{2}$, J.A.~Hinton$^{1,3}$,
  Nu.~Komin$^{2}$, K.~Kosack$^{1}$ and C.~Stegmann$^{4}$ for the
  H.E.S.S. collaboration}
\shortauthors{S. Carrigan et al.}
\afiliations{
\small
  $^1$Max-Planck-Institut f\"ur Kernphysik, P.O. Box 103980, D 69029 Heidelberg, Germany  \\
  $^2$Laboratoire de Physique Th\'eorique et Astroparticules, IN2P3/CNRS, Universit\'e Montpellier II, CC 70, Place Eug\`ene Bataillon, F-34095 Montpellier Cedex 5, France  \\
  $^3$Landessternwarte, Universit\"at Heidelberg, K\"onigstuhl, D 69117 Heidelberg, Germany  \\
  $^4$Universit\"at Erlangen-N\"urnberg, Physikalisches Institut, Erwin-Rommel-Str. 1, D 91058 Erlangen, Germany}
\email{svenja.carrigan@mpi-hd.mpg.de}

\abstract{Motivated by recent detections of pulsar wind nebulae in
  very-high-energy (VHE) gamma rays, a systematic search for VHE
  gamma-ray sources associated with energetic pulsars was performed,
  using data obtained with the H.E.S.S. (High Energy Stereoscopic
  System) instrument. The search for VHE gamma-ray sources near the
  pulsar PSR J1718-3825 revealed the new VHE gamma-ray source HESS
  J1718-385. We report on the results from the HESS data analysis of
  this source and on possible associations with the pulsar and at
  other wavelengths. We investigate the energy spectrum of HESS
  J1718-385 that shows a clear peak. This is only the second time a
  VHE gamma-ray spectral maximum from a cosmic source was observed,
  the first being the Vela X pulsar wind nebula.}

\begin{document}
\maketitle

\section{Introduction}

It has long been known that pulsars can drive powerful winds of highly
relativistic particles. Confinement of these winds leads to the
formation of strong shocks, which may accelerate particles to
$\sim$PeV energies.

The best studied example of a pulsar wind nebula (PWN) is the Crab
nebula, which exhibits strong non-thermal emission across most of the
electromagnetic spectrum from radio to $>$50~TeV $\gamma$-rays
\cite{WHIPPLE:crab}.  More recently, VHE $\gamma$-ray emission has
been detected from the Vela\,X PWN \cite{HESS:velax}, which is an
order of magnitude older ($\sim$11\,kyr) than the Crab nebula, and its
nebula is significantly offset from the pulsar position, both in
X-rays and VHE $\gamma$-rays.  Offset nebulae in both X-rays and VHE
$\gamma$-rays have also been observed in the Kookaburra Complex
\cite{HESS:kookaburra} and for the PWN associated with the
$\gamma$-ray source HESS~J1825$-$137 \cite{HESS:J1825a,HESS:J1825b}.
The latter source appears much brighter and more extended in VHE
$\gamma$-rays than in keV X-rays.  This suggests that searches at TeV
energies are a powerful tool for detecting PWNe.

Motivated by these detections, a systematic search for VHE
$\gamma$-ray sources associated with high spin-down energy loss rate
pulsars was performed, using data obtained with the
H.E.S.S. instrument. The VHE $\gamma$-ray data set used in the search
includes all data used in the H.E.S.S. Galactic plane survey
\cite{HESS:scanpaper2}, an extension of the survey to $-60^{\circ} < l
< -30^{\circ}$, dedicated observations of Galactic targets and
re-observations of H.E.S.S. survey sources. It spans Galactic
longitudes $-60^{\circ} < l < 30^{\circ}$ and Galactic latitudes
$-2^{\circ} < b < 2^{\circ}$, a region covered with high sensitivity
in the survey. These data are being searched for VHE emission from
pulsars from the Parkes Multibeam Pulsar Survey~\cite{Parkes1}. The
search for a possible $\gamma$-ray excess is done in a circular region
with radius $\theta = 0.22^{\circ}$ (as in \cite{HESS:scanpaper2})
around each pulsar position, sufficient to encompass a large fraction
of a possible PWN.  The statistical significance of the resulting
associations of the VHE $\gamma$-ray source with the pulsar is
evaluated by repeating the procedure for randomly generated pulsar
samples, modelled after the above-mentioned parent population.

In this search, it is found that pulsars with high spin-down energy
loss rates are on a statistical basis accompanied by VHE emission.
The search for VHE $\gamma$-ray emission near the pulsar \PSRONE\
revealed the new VHE $\gamma$-ray source \HESSONE. This paper deals
with the results from the HESS data analysis of \HESSONE\ and with its
possible associations with \PSRONE\ and other objects seen in radio
and X-ray wavelengths.

\section{H.E.S.S. Observations and Analysis}

The data on \HESSONE\ are composed primarily from dedicated
observations of the supernova remnant RX~J1713.7$-$3946
\cite{HESS:RXJ1713}, which is located at about 1.6$^\circ$ south-west
of \HESSONE. After passing the H.E.S.S.\ standard data quality
criteria based on hardware and weather conditions, the data set for
\HESSONE\ has a total live time of $\sim$82 hours. The standard
H.E.S.S.\ analysis scheme \cite{HESS:crab} is applied to the data,
including optical efficiency corrections. In this analysis, \emph{hard
  cuts} are applied, which include a rather tight cut on the shower
image brightness of 200 photo-electrons and are suitable for extended,
hard-spectrum sources such as PWN. These cuts also improve the angular
resolution and therefore suppress contamination from the nearby
RX~J1713.7$-$3946. To produce a sky map, the background at each test
position in the sky is derived from a ring surrounding this position
with a mean radius of 1$^\circ$ and a width scaled to provide a
background area that is about 7 times larger than the area of the
on-source region.

For spectral studies, only observations in which the camera centre is
offset by less than 2$^\circ$ from the best-fit source position are
used to reduce systematic effects due to reconstructed $\gamma$-ray
directions falling close to edge of the field of view. The remaining
live time of the data sample is $\sim$73 hours. The spectral
significance is calculated by counting events within a circle of
radius 0.2$^\circ$ from the best-fit position, chosen to enclose the
whole emission region while reducing systematic effects arising from
morphology assumptions. The proximity of the strong source makes it
necessary to choose the background data from off-source observations
(matched to the zenith angle and offset distribution of the on-source
data) instead of from areas in the same field of view. For a more
detailed description of methods for background estimation, see
\cite{HESS:bg}.

\section{Results}

The detection significance from the search for VHE $\gamma$-ray
emission within 0.22$^\circ$ of the location of \PSRONE\ is
$7.9\sigma$. A very conservative estimate of the number of trials
involved (\cite{HESS:scanpaper2}) leads to a corrected significance of
$6.2\sigma$.

\begin{figure}[!hbt!]
  \centering
  \includegraphics[width=0.48\textwidth]{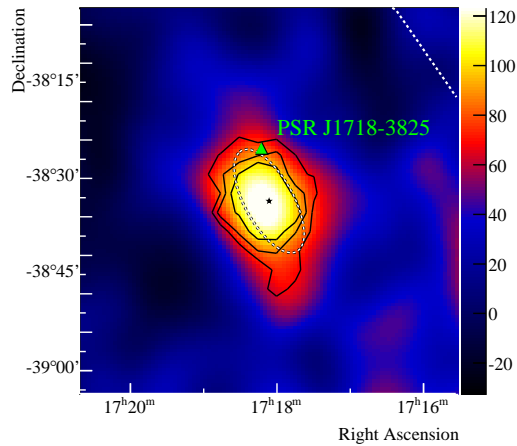}
  \caption{\label{PP_1718_figA}\small An image of the VHE $\gamma$-ray
    excess counts of \HESSONE, smoothed with a Gaussian of width
    0.06$^\circ$. The colour scale is set such that the blue/red
    transition occurs at approximately the 3$\sigma$ significance
    level. The black contours are the 4, 5 and 6$\sigma$ significance
    contours. The position of the pulsar \PSRONE\ is marked with a
    green triangle and the Galactic plane is shown as a white dotted
    line.  The best-fit position for the $\gamma$-ray source is marked
    with a black star and the fit ellipse with a dashed line.}
\end{figure}
  
Figure \ref{PP_1718_figA} shows the smoothed excess count map of the
1$^\circ$~$\times$~1$^\circ$ region around \HESSONE. A two-dimensional
Gaussian brightness profile, folded with the H.E.S.S. point-spread
function, is fit to the distribution before smoothing. Its parameters
are the width in two dimensions and the orientation angle, defined
counter-clockwise from North. The intrinsic widths (with the effect of
the point-spread function removed) for the fit are $9' \pm 2'$ and $4'
\pm 1'$ and the orientation angle is $\sim$33$^{\circ}$. The best-fit
position for the centre of the excess is RA~=~\HMS{17}{18}{7}\,$\pm
5^{s}$, Dec~=~$-38^{\circ}33'\pm2'$ (epoch J2000). H.E.S.S. has a
systematic pointing error of $\sim 20''$.

\begin{figure}[!hbt!]
  \centering
  \vspace{3ex}
  \includegraphics[width=0.48\textwidth]{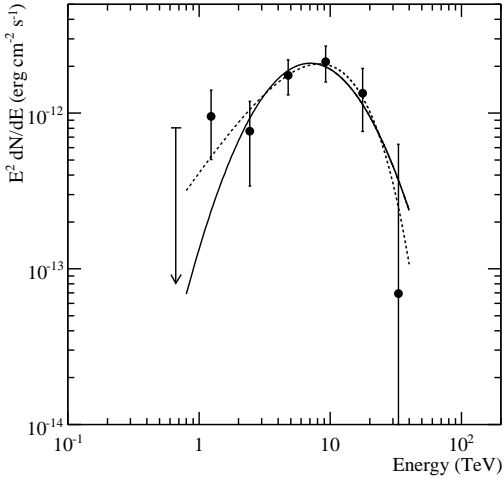}
  \caption{\label{J1718-both}\small The energy spectrum of \HESSONE, which is
    fit by a curved profile (solid line).  Alternatively, the fit of
    an exponentially cut-off power law is shown (dashed line, refer to
    the text for details on both fits). The first point in the
    spectrum lacks statistics due to lower exposure at small zenith
    angles and is plotted as an upper limit with at a confidence level
    of $2\sigma$.}
\end{figure}

For the spectral analysis, a statistical significance of $6.8\sigma$
(with 343 excess counts) is derived. Figure \ref{J1718-both} shows the
measured spectral energy distribution for \HESSONE\ (in $E^{2}\,dN/dE$
representation).

The spectrum is fit by a curved profile (shown as the solid line):
\begin{equation}
  \frac{dN}{dE} = N_0 \left(\frac{E_{\mathrm{peak}}}{1\,\mathrm{TeV}}\right)^{-2} \left(\frac{E}{E_{\mathrm{peak}}}\right)^{\beta \cdot \mathrm{ln}(E/E_{\mathrm{peak}}) - 2}
\end{equation}
The peak energy $E_{\mathrm{peak}}$ is $(7 \pm 1_\mathrm{stat} \pm
1_\mathrm{sys})$\,TeV, the differential flux normalisation $N_0 = (1.3
\pm 0.3_\mathrm{stat} \pm 0.5_\mathrm{sys}) \times
10^{-12}$\,TeV$^{-1}$\,cm$^{-2}$\,s$^{-1}$ and $\beta = -0.7 \pm
0.3_\mathrm{stat} \pm 0.4_\mathrm{sys}$. This fit has a
$\chi^2/{d.o.f.}$ of $3.2/3$. The integral flux between $1-10$~TeV is
about 2\,\% of the flux of the Crab nebula in the same energy
range~\cite{HESS:crab}. 

Alternatively, fitting the spectrum by an exponentially cut-off power
law ($dN/dE = N_0 E^{-\Gamma} e^{-E/E_{\mathrm{cut}}}$) gives $N_0 =
(3.0 \pm 1.9_\mathrm{stat} \pm 0.9_\mathrm{sys}) \times
10^{-13}$\,TeV$^{-1}$\,cm$^{-2}$\,s$^{-1}$, photon index $\Gamma = 0.7
\pm 0.6_\mathrm{stat}\pm 0.2_\mathrm{sys}$ and a cut-off in the
spectrum at an energy of $E_{\mathrm{cut}}=(6 \pm 3_\mathrm{stat} \pm
1_\mathrm{sys})$\,TeV. This fit, which is shown as a dashed line in
Figure \ref{J1718-both}, has a $\chi^2/{d.o.f.}$ of $1.6/3$.

Both the curved and exponentially cut-off power law profiles fit the
data well; the former has the advantage of showing explicitly the
peak energy of the spectrum, which has to date only been resolved in
one other VHE source, Vela\,X \cite{HESS:velax}.

\section{Possible Associations}

The $\gamma$-ray source \HESSONE\ is located $\sim$0.14$^\circ$ south
of the pulsar \PSRONE. \PSRONE\ appears to be a Vela-like pulsar, as
it is of comparable age, 90~kyr, and has a similar spin period,
75\,ms.  From the spectral fit of a curved profile, the energy flux of
\HESSONE\ between (1 -- 10)\,TeV is estimated to
$2.9\times10^{-12}$~erg cm$^{-2}$s$^{-1}$.  With a distance of
$\sim$4\,kpc and a spin-down luminosity of $\dot{E} =
1.3\times10^{36}$\,erg\,s$^{-1}$, \PSRONE\ is energetic enough to
power \HESSONE, with an implied efficiency of $\epsilon_\gamma \equiv
L_\gamma/\dot{E} = 0.5\,\%$.

\begin{figure}[!hbt!]
  \centering
  \vspace{3ex}
  \includegraphics[width=0.48\textwidth]{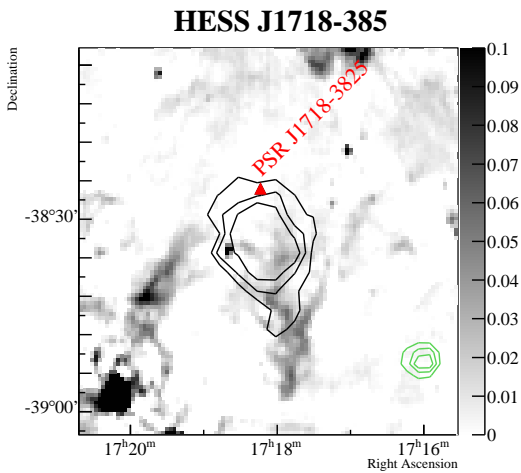}
  \caption{\label{PP_1718_figB}\small Radio image from the Molonglo Galactic
    Plane Survey at 843\,MHz~\cite{Molonglo} (in Jy/beam). The
    H.E.S.S. significance contours are overlaid in black and the
    pulsar position is marked with a red triangle. Adaptively smoothed
    ROSAT hard-band X-ray contours are shown in green~\cite{ROSAT}.}
\end{figure}

As can be seen in Figure \ref{PP_1718_figB}, no obvious X-ray
counterpart is visible for \HESSONE. There is diffuse extended radio
emission, which is partially coincident with the VHE emission.
However, this emission seems to be correlated with thermal dust
emission visible in the IRAS Sky Survey Atlas~\cite{IRAS}, suggesting
that the radio emission is thermal and is thus not likely associated
with a possible PWN. The brightest part of this diffuse feature is
catalogued as PMN~J1717$-$3846~\cite{PMN:1}. From the point of view of
positional coincidence, energetics, and lack of other counterparts,
the association of \HESSONE\ with \PSRONE\ seems plausible. To confirm
this, additional evidence from spectral and morphological studies in
VHE $\gamma$-rays and from data at other wavelengths is needed.

\HESSONE\ may well represent the first VHE $\gamma$-ray PWN found in a
systematic search for pulsar associations, despite the present lack of
a PWN detection in other wave bands. The remarkable similarity between
\HESSONE\ and other known VHE PWNe, together with the lack of other
probable counterparts, gives additional confidence. The detection of
an X-ray PWN would provide confirmation.

\section{Acknowledgements}
\small The support of the Namibian authorities and of the University
of Namibia in facilitating the construction and operation of H.E.S.S.
is gratefully acknowledged, as is the support by the German Ministry
for Education and Research (BMBF), the Max Planck Society, the French
Ministry for Research, the CNRS-IN2P3 and the Astroparticle
Interdisciplinary Programme of the CNRS, the U.K. Science and
Technology Facilities Council (STFC), the IPNP of the Charles
University, the Polish Ministry of Science and Higher Education, the
South African Department of Science and Technology and National
Research Foundation, and by the University of Namibia. We appreciate
the excellent work of the technical support staff in Berlin, Durham,
Hamburg, Heidelberg, Palaiseau, Paris, Saclay, and in Namibia in the
construction and operation of the equipment.\\We have made use of the
ROSAT Data Archive of the Max-Planck-Institut fuer extraterrestrische
Physik (MPE) at Garching, Germany.

\bibliography{icrc0494}
\bibliographystyle{plain}

\end{document}